\documentclass[moor,sglanonrev]{informs4_arxiv}
\usepackage{eqndefns-left} 
\RequirePackage{tgtermes}
\RequirePackage{newtxtext}
\RequirePackage{newtxmath}
\RequirePackage{bm}
\RequirePackage{endnotes}

\usepackage{tikz}
\usetikzlibrary{automata, positioning, arrows}
\OneAndAHalfSpacedXII 

\usepackage{algorithm}
\usepackage{algpseudocode}
\usepackage{tikz}

\EquationsNumberedThrough    

\TheoremsNumberedThrough     
\ECRepeatTheorems  %

\MANUSCRIPTNO{MOOR-0001-2024.00}

\usepackage{amsmath}

\newcommand{\jgc}[1]{\textcolor{blue}{[jgc: #1]}}

\newcommand{\calP}{{\cal P}}

\newcommand{\ext}{\texttt{ext}}
\newcommand{\ep}{\varepsilon}
\newcommand{\dN}{{{\bf N}}}

\newcommand{\dR}{{{\bf R}}}
\newcommand{\E}{{{\bf E}}}

\newcommand{\prob}{{{\bf P}}}

\newcommand{\X}{\mathcal{S}}
\newcommand{\M}{\mathcal{M}}

\newcommand{\cR}{\mathcal{R}}
\newcommand{\A}{\mathcal{A}}
\newcommand{\U}{\mathcal{U}}

\usepackage[colorinlistoftodos,bordercolor=orange,backgroundcolor=orange!20,linecolor=orange,textsize=scriptsize]{todonotes}

\numberwithin{equation}{section}
\numberwithin{theorem}{section}
\numberwithin{proposition}{section}

\begin{document}


\RUNAUTHOR{Grand-Cl{\'e}ment and Vieille}

\RUNTITLE{Playing against a stationary opponent}

\TITLE{Playing against a stationary opponent}

\ARTICLEAUTHORS{%

\AUTHOR{Julien Grand-Cl{\'e}ment}
\AFF{Department of Information Systems and Operations Management,
HEC Paris, \EMAIL{grand-clement@hec.fr}}

\AUTHOR{Nicolas Vieille}
\AFF{Department of Economics and Decision Sciences,
HEC Paris, \EMAIL{vieille@hec.fr}}

} 

\ABSTRACT{%
This paper investigates properties of Blackwell $\ep$-optimal strategies in zero-sum stochastic games when the adversary is restricted to stationary strategies, motivated by applications to robust Markov decision processes. For a class of absorbing games, we show that Markovian Blackwell $\ep$-optimal strategies may fail to exist, yet we prove the existence of Blackwell $\ep$-optimal strategies that can be implemented by a two-state automaton whose internal transitions are independent of actions. For more general absorbing games, however, there need not exist Blackwell $\ep$-optimal strategies that are independent of the adversary's decisions. Our findings point to a contrast between absorbing games and generalized Big Match games, and provide new insights into the properties of optimal policies for robust Markov decision processes. 
}





\maketitle

\section{Introduction}




The purpose of this short paper is to solve a non-conventional question on zero-sum  stochastic games that is motivated by robust Markov decision processes (MDPs). Alternatively, our objective is to analyze a question relative to robust MDPs by means of techniques from stochastic games.

Robust MDPs are a framework for the formal analysis of Markov decision processes in which the decision maker (DM) is unsure of the transition function, see \cite{nilim2005robust,iyengar2005robust,wiesemann2013robust}. In line with a tradition that dates back to Wald \cite{wald1949statistical}, this uncertainty is modeled by assuming that `nature' reacts adversarially to the DM's strategy by choosing a transition function that minimizes the DM's payoff. 
As such, it is clear that robust MDPs and zero-sum stochastic games are strongly related, even if there is no one-to-one mapping between the questions addressed in the two communities. The model of {\em $s$-rectangular} robust MDPs~\cite{le2007robust,wiesemann2013robust} relates to `canonical' zero-sum stochastic games, while $sa$-rectangular robust MDPs~\cite{nilim2005robust,iyengar2005robust} correspond instead to perfect information stochastic games, and non-rectangular robust MDPs~\cite{goyal2023robust} have no analog in stochastic games. 

A critical difference is that in robust MDPs the transition function is chosen adversarially once and for all, see \cite{grand2023beyond}. In a stochastic game, the second player (P2) instead faces no such constraint, which  impacts the structure of the optimal strategies for the first player (P1). 

An emerging literature focuses on Blackwell $\ep$-optimal policies in robust MDPs, i.e. on policies that are $\ep$-optimal (against stationary strategies) in all discounted problems with a discount rate sufficiently close to $0$, see \emph{e.g.}, \cite{grand2023beyond,wang2023robust}. In \cite{grand2023beyond}, it is shown for $sa$-rectangular robust MDPs (perfect information stochastic games) that there always exists a stationary deterministic policy that is Blackwell $\ep$-optimal for all $\ep>0$. 
Blackwell (0-)optimal policies  exist under an additional {\em definability} assumption.
In \cite{wang2023robust}, it is shown that such Blackwell (0)-optimal policies exist for \emph{unichain} sa-rectangular problems under the additional assumption that the average optimal policy is unique. Here, we aim to find results for $s$-rectangular problems/stochastic games. 

In arbitrary finite stochastic games, the existence of Blackwell $\ep$-optimal strategies follows from the existence of a uniform value~\cite{mertens1982stochastic}. Yet, such strategies depend on the opponent's past play in a typically complex way. In fact, in a game as innocent-looking as the Big Match~\cite{gillette,blackwell1968big}, no Markovian strategy is Blackwell $\ep$-optimal, nor is any strategy of P1 that can be implemented by some arbitrary finite automaton. In other words, any such strategy of P1 can be defeated by a strategy of P2, as long as P2 is not restricted to stationary strategies.

Our main question in this paper is the following. How simple can Blackwell $\ep$-optimal strategies of P1 be, \emph{assuming} that P2 is restricted to choose a stationary strategy?  The question is, therefore, to identify 'simple' strategies that approximately guarantee the discounted value against any stationary opponent, provided that the discount rate is small enough. We address this question via two classes of stochastic games.  \emph{Absorbing games} are instances in which the state changes at most once over time. 
While simpler to analyze than general stochastic games, absorbing games have proven a useful testbed to get insights relevant more generally, see \textit{e.g.}~\cite{kohlberg1974repeated,vrieze1989equilibria,solan1999three,mertens2009absorbing,flesch1997cyclic}. Among such  games, the  Big Match games and their variants, such as the \textit{generalized Big Match} games~\cite{coulomb1992repeated}, play a prominent role. It seems a consensus view that these games are qualitatively no easier to play than absorbing games. 

Our results are mixed. Our main positive result applies to a (strict) subclass of absorbing games that includes all generalized Big Match games. Within this class of games (and for each $\ep>0$), P1 has a Blackwell $\ep$-optimal strategy that can be implemented by a 2-state automaton, whose transitions are independent of past action choices. For such strategies, P1 switches between two mixed actions at random times, independently of the entire past play. Surprisingly however, \emph{Markovian} Blackwell $\ep$-optimal strategies may fail to exist.
On the other hand,  this result does not extend to all absorbing games, and we provide a game in which P1 has no Blackwell $\ep$-optimal strategy that does not depend on P2's actions. 

The implications are twofold: (i) when P2 is constrained to stationary strategies, generalized Big Match games are much easier to play than absorbing games; (ii) except in specific classes of games, facing a stationary opponent does not enable P1 to dispense with the complexity of history-dependent strategies. 

\paragraph{Outline.}
The paper is organized as follows. Section \ref{sec setup} contains the formal setup and the statement of the main results. In Section \ref{sec examples}, we analyze a variant of the Big Match, in which P1 has a simple strategy that is Blackwell $\ep$-optimal for each $\ep>0$ when P2 is restricted to stationary strategies, yet no {\em Markovian} strategy is Blackwell $\ep-$optimal. The latter claim relies on the observation that geometrically distributed random variables (r.v.'s) cannot be approximated by a possibly infinite sum of independent Bernoulli r.v.'s with arbitrary parameters. We leverage this analysis in a class of absorbing games that includes the generalized Big Matches in Section \ref{sec GBM}. Section \ref{sec general} discusses an absorbing game that falls outside of this class. We use the language and formalism of stochastic games throughout this paper, and we discuss the implications of our results for robust MDPs in Section \ref{sec RMDPs}.

\section{Setup and main results}\label{sec setup}

\subsection{Setup}
A zero-sum stochastic game (SG) $\Gamma$ is described by a set of states $S$, an action set $A$ and $B$ for each of two players P1 and P2, a reward function $r:S\times A\times B\to \dR$ and a transition function $p: S\times A\times B\to \Delta(S)$. The sets $S$, $A$ and $B$ are assumed to be finite. Given a finite set $K$, $\Delta(K)$ denotes the simplex of probability distributions over $K$.

The game is played in stages indexed by $t \in \dN = \{1,2,...\}$. In each stage $t$, the game is in some state $s_t\in S$ and the two players choose actions $a_t\in A$ and $b_t\in B$. 
We assume that  $s_t$ is known to the players when they choose actions, and that actions are chosen simultaneously and publicly. The reward in stage $t$ is $r_t:= r(s_t,a_t, b_t)$ and the next state $s_{t+1}$ is drawn according to the distribution $ p(\cdot \mid s_t,a_t, b_t)$. 

For $t\geq 1$, we denote by $H_t:= (S\times A\times B)^{t-1}\times S$ the set of histories up to stage $t$. The information available in stage $t\geq 1$ is the sequence $h_t= (s_1,a_1,b_1,\cdots, b_{t-1}, s_t)\in H_t$ of past states and actions together with the current state $s_t$. A (behavior) strategy of P1 (resp., of P2) is therefore a function $\sigma: \cup_{t\geq 1}H_t \to \Delta(A)$ (resp. $\tau:\cup_{t\geq 1} H_t \to \Delta(B)$). 
A strategy $\sigma$ is \emph{stationary} if $\sigma(h_t)$ only depends on $s_t$. A stationary strategy of P1 (resp. of P2) is identified with a vector $x\in \Delta(A)^S$ (resp., $y\in \Delta(B)^S$) with the interpretation that P1's action is drawn using $x_s\in \Delta(A)$ whenever the current state is $s$. 

Given a state $s\in S$ and a strategy profile $(\sigma,\tau)$, we denote by $\prob_{s,\sigma,\tau}$ the probability distribution over the set  $H_\infty= (S\times A\times B)^{\dN}$ of infinite plays when the game starts from $s_1= s$ and players behave according to $\sigma$ and $\tau$. Expectations under $\prob_{s,\sigma,\tau}$ are denoted $\E_{s,\sigma,\tau}$. 
Given a discount rate $\lambda >0$, the induced (normalized) $\lambda$-discounted payoff is 
\begin{equation}\label{eq:discount payoff}
\gamma_\lambda(s,\sigma,\tau):=\E_{s,\sigma,\tau}\left[\lambda\sum_{t= 1}^{+\infty} (1-\lambda)^{t-1} r_t\right]
\end{equation}

\subsection{Definitions}

For $s\in S$,  the zero-sum game with payoff function $\gamma_\lambda(s,\cdot, \cdot)$ has a value $v_\lambda (s)$ and  optimal stationary strategies $x$ and $y$, see \cite{shapley1953stochastic}. That is, 
\[v_\lambda(s)= \inf_{\tau} \gamma_\lambda(s,x,\tau)= \sup_{\sigma} \gamma_\lambda(s,\sigma,y).\]

Blackwell ($\ep$-)optimal strategies are strategies that are ($\ep$-)optimal irrespective of the value of $\lambda$, provided $\lambda$ is small enough. Given that our motivation relates to robust MDPs where the adversary is stationary, we use the following definition.
\begin{definition}
    Let $\ep>0$. A strategy $\sigma$ is {\em Blackwell $\ep$-optimal } if there exists $\lambda_0>0$ such that
    \[\gamma_\lambda(s,\sigma,y)\geq v_\lambda(s)-\ep,\]
    for each $0<\lambda<\lambda_0$, $y\in \Delta(B)^S$ and $s\in S$.
\end{definition}
We emphasize that in this definition, the optimality properties of $\sigma$ are only tested against \emph{stationary} strategies $y\in \Delta(B)^S$ of P2.
\medskip

An \emph{absorbing state} is a state $s^*\in S$ such that $p(s^*\mid s^*,a,b)=1$ for each $(a,b)\in A\times B$. An \emph{absorbing game} is a game where all states are absorbing except one. 

We focus on such games and we decompose the state space as $S=S^*\cup\{s\}$, where $S^*$ is the subset of absorbing states, and $s$ is the (only) non-absorbing state. 
For all relevant purposes, it is without loss of generality to assume that the rewards in an absorbing state are independent of action choices, and we simply write $r(s^*)$ for the payoff in state $s^*\in S^*$. We also write $r(a,b)$ and $p(\cdot \mid a,b)$ instead of $r(s,a,b)$ and $p(\cdot \mid s,a,b)$, and we denote by  $p^*(a,b):=p(S^*\mid a,b)$ the probability of moving to an absorbing state when playing $(a,b)$.

\begin{definition}
    We say that the stochastic game $\Gamma= (S,A,B,p,r)$ is a \emph{product absorbing} game if it is absorbing and if there exist $A^*\subseteq A$ and $B^*\subseteq B$ such that 
    \[p^*(a,b)>0 \Leftrightarrow (a,b)\in A^*\times B^*.\]
\end{definition}

Up to a reordering of the actions,  probabilities $p^*(a,b)$ of termination then have the following block structure (as a matrix indexed by $(a,b) \in A \times B$):

\[
\bordermatrix{
         & B^*   &  B\setminus B^*  \cr
      A^*  & >0 & 0  \cr
      A\setminus A^*  & 0   &  0  \cr
}
\]

Generalized big match games, see \cite{coulomb1992repeated}, are product absorbing games such that $B^*=B$ and $|A|=2$.

As pointed out in the introduction, the existence of Blackwell $\ep$-optimal strategies follows from the existence of a uniform value~\cite{mertens1982stochastic}, but these strategies may be complex (history-dependent). Our contribution relates to the existence or non-existence of \emph{simple} Blackwell $\ep$-optimal strategies. We introduce three such classes of strategies. 

A strategy $\sigma$ is \emph{Markovian} if $\sigma(h_t) \in \Delta(A)$ only depends on $t$ and $s_t$, for any stage $t\geq 1$ and each history $h_t \in H_t$. In an absorbing game, a Markovian strategy can  be identified with the sequence $(x_t)_{t\geq 0}$, where $x_t\in \Delta(A)$ is the (mixed) action of P1 in stage $t$, if $s_t=s$. 

A strategy $\sigma$ is \emph{blind} if $\sigma(h_t)\in \Delta(A)$ is independent of the past actions of P2.
Blind strategies may depend on P1's  own past actions. 

An \emph{automaton} (see \cite{neyman1998finitely}) is a strategy with the property that the (mixed) action of P1 in any stage only depends on the internal state of the automaton, which is updated over time.\footnote{See \cite{kalai1988finite} for a discussion of strategy complexity, in the context of repeated games.} Formally, a (finite) \emph{automaton} is a tuple $(K,\mu_0,\pi,f)$, where $K$ is a finite set of internal states, $\mu_0\in \Delta(K)$ is the distribution of the initial internal state, and $\pi: K\times A\times B\to \Delta(K)$ is the transition rule of the automaton, with the interpretation that $\pi(\cdot \mid k,a,b)$ is the distribution of the next internal state, as a function of the current internal state, and of the actions currently chosen by both players. Finally, $f:K\to \Delta(A)$ specifies P1's action choice, as a function of the current internal state of the automaton. The \emph{size} of the automaton is the cardinality of $K$. 
We say that the automaton is \emph{autonomous} if the transitions $\pi(\cdot \mid k,a,b)$ are independent of the actions $a,b$ of the players. An autonomous automaton thus randomly switches between finitely many mixed actions $f(k)$, $k\in K$, independently of past play. We provide an example in Section \ref{proof first}.
We note that Markovian strategies and autonomous automata are blind, but that the sets of Markovian strategies and of automata are non-nested.

\subsection{Main results}
Our main results are Theorem \ref{th main} and Proposition \ref{prop counter} below.
\begin{theorem}\label{th main}
The following results hold:
\begin{itemize}
    \item There is a product absorbing game and $\ep>0$, such that no Markovian strategy of P1 is Blackwell $\ep$-optimal.
    \item For each product absorbing game and each $\ep>0$, there is a Blackwell $\ep$-optimal strategy that can be implemented as an autonomous automaton of size 2.
\end{itemize}
\end{theorem}

 The automata we construct share the additional property that one of the two internal states is absorbing. For such an automaton, the behavior of  P1 is characterized by (at most) two regimes. P1 first repeats the same mixed action for a random duration $D$, 
 then switches to another mixed action forever when the automaton transitions to the absorbing internal state.  The duration $D$ of the first phase follows a geometric distribution. In our view, such strategies, while not Markovian, are arguably even simpler.

Our second main result is the following proposition.
\begin{proposition}\label{prop counter}
    There is an absorbing game and $\ep>0$ such that P1 has no blind Blackwell $\ep$-optimal strategy.
\end{proposition}

Together, Theorem \ref{th main} and Proposition \ref{prop counter} suggest that the restriction on P2's behavior leads to a drastic simplification of Blackwell $\ep$-optimal strategies in product absorbing games, but not beyond. 

\section{Blackwell optimality in product absorbing games}\label{sec examples}
We focus in this  section on product absorbing games.  We first revisit the classical  Big Match game, assuming that  P2 is restricted to stationary strategies. For product absorbing games, we then prove the suboptimality of Markovian strategies (first statement of Theorem \ref{th main}) in Section \ref{sec modified}, and the optimality of autonomous automaton (second statement of Theorem \ref{th main}) in Section \ref{sec GBM}.
\subsection{The Big Match}\label{sec: big match}
Table 1 below describes a product absorbing game known as the Big Match, introduced in \cite{gillette} and first analyzed in \cite{blackwell1968big}. The entries of the table contain the payoff of P1 as a function of $(a,b)$ in the non-absorbing state. Rows (\texttt{Top} and \texttt{Bottom}) and columns (\texttt{Left} and \texttt{Right}) are the actions  of P1 and P2 respectively. Starred entries encode transitions: as long as P1 plays the bottom row, the game stays in the initial, non-absorbing state; as soon as P1 plays the top row, the game moves to an absorbing state, with a payoff of either 0 or 1. 

\[ \begin{array}{|c|c|}
   \hline
    1^* & 0^* \\
    \hline
    0 & 1 \\
    \hline
    \end{array}\]
\begin{center}
Table 1: The Big Match 
\end{center}

 When P2 is unrestricted, it is known that for $\ep>0$ small enough, there is no Blackwell $\ep$-optimal strategy of P1 that is either Markovian or that can be implemented by  some finite automaton, see \cite{MSZ}, p. 417.
On the other hand, the strategy $\sigma$ of P1 that plays the two rows with equal probabilities in the first stage, then repeatedly plays the bottom row, satisfies $\gamma_\lambda(\sigma,y)=\frac12$ for each $\lambda$ and $y\in \Delta(B)$. Since $v_\lambda = \frac12$ for each $\lambda>0$, $\sigma$ is a Blackwell 0-optimal strategy. Note that $\sigma$  is Markovian and that it can also be implemented as an autonomous automaton with two states.

\subsection{A modified Big Match}\label{sec modified}

 In this subsection, we analyze the variant of the Big Match presented in Table 2.
\[ \begin{array}{|c|c|c|}
   \hline
    1^* & 0^* & \frac12\\
    \hline
    0 & 1 & \frac12\\
    \hline
    \end{array}
    \]
\begin{center}
Table 2: A Modified Big Match 
\end{center}

We denote the three actions of P2 as \texttt{Left, Middle} and \texttt{Right} and the two actions of P1 as \texttt{Top} and \texttt{ Bottom}. This game is obtained from the Big Match by adding one action for P2 in the initial state, hence its discounted value $v_\lambda$ cannot exceed the discounted value of the Big Match: $v_\lambda\leq \frac12$. On the other hand, denoting $x_\lambda\in \Delta(\{\texttt{Top,Bottom}\})$ the optimal strategy of P1 in the Big Match,\footnote{$x_\lambda$ assigns probability $\frac{\lambda}{1+\lambda}$ to $\texttt{Top}$} it is immediate that $\gamma_\lambda(x_\lambda,y)= \frac12$ for each $y\in \Delta(B)$, hence $v_\lambda= \frac12$ and $x_\lambda$ is optimal in the modified Big Match as well.

We will prove Proposition \ref{prop MBM} below. Note that the second statement coincides with the first statement of Theorem \ref{th main}.

\begin{proposition}\label{prop MBM}
    For the modified Big Match, the following holds.
    \begin{itemize}
        \item There exists a  Blackwell 0-optimal strategy that can be implemented by an autonomous automaton of size 2.
        \item For $\ep>0$ small enough, there is no Markovian  Blackwell $\ep$-optimal strategy.
    \end{itemize}
\end{proposition}
We prove the first part in Section \ref{proof first} and the second part in Section \ref{proof second}. 

\subsubsection{On Blackwell $\ep$-optimal strategies in the Modified Big Match}\label{proof first}

Here and elsewhere, we write $\tau:=\inf\{t: s_{t}\in S^*\}$ for the absorption stage (with $\inf\emptyset = 0$). Even though the game ends in stage $\tau$, it is convenient to assume that players keep selecting actions for ever after.

We denote by $N$ the random number of times P1 plays $\texttt{Top}$ over time: 
\[N:=\#\{t\geq 1: a_t= \texttt{Top}\}.\] 
Given a blind strategy $\sigma$, the distribution of $N$ is independent of the stationary strategy $y$ of P2, and is denoted
 $\prob_\sigma\in \Delta(\dN\cup \{0,+\infty\})$. Expectations with respect to $\prob_\sigma$ are written $\E_\sigma$.

Given a stationary strategy  $y\in \Delta(B)$ of P2, we denote by $y(L), y(M)$ and $y(R)$ the probabilities assigned to the actions \texttt{Left}, \texttt{Middle} and \texttt{Right} respectively.

\begin{lemma}\label{lemm1} Let $\sigma$ be a blind strategy of P1, and $y\in \Delta(B)$. 
\begin{itemize}
    \item We have
    \begin{equation}\label{eq1}\lim_{\lambda \to 0} \gamma_\lambda(\sigma,y) = \left(1-
\E_\sigma\left[y(R)^N\right]\right) \frac{y(L)}{y(L)+y(M)} + \E_\sigma\left[y(R)^N\right]\left(y(M) +\frac12 y(R)\right). \end{equation}
\item If $\E_\sigma[N]<+\infty$, the convergence in (\ref{eq1}) is uniform over $y\in \Delta(B)$.
\end{itemize}

\end{lemma}

We set $\gamma(\sigma,y):=\lim_{\lambda \to 0} \gamma_\lambda(\sigma,y)$.
\medskip
\proof{Proof.}
If $y(R)=1$, one has $\gamma_\lambda(\sigma,y)= \frac12$ for each $\lambda>0$, so the conclusion holds. Assume now that $y(R)<1$ and denote by  $\displaystyle \bar r_\lambda:= \lambda \sum_{t=1}^{+\infty}(1-\lambda)^{t-1}r_t$ the realized $\lambda$-discounted payoff.

Whenever $a_t=\texttt{Top}$ (and if $\tau> t$), the game moves to $S^*$ with probability $y(L) +y(M)$, and the conditional probability of ending in state $1^*$ is $\frac{y(L)}{y(L)+y(M)}$. The former observation implies that
 $\prob_{\sigma,y}\left(\tau=+\infty \mid N\right)= y(R)^N$. The latter one implies that conditional on $\tau<+\infty$, $\bar r_\lambda$ converges to 1 with probability $\frac{y(L)}{y(L)+y(M)}$ and to 0 otherwise.

On the other hand, the successive actions $(b_t)_{t \in \dN}$ of P2 are independent and identically distributed according to $y\in \Delta(B)$, hence the empirical distribution of $b_t$ converges $\prob_{\sigma,y}$-almost surely (a.s.) to $y$. It follows that $\lim_{\lambda\to 0} \bar r_\lambda= y(M) +\frac12 y(R)$ a.s. on the event $\tau = +\infty$.
By dominated convergence, this implies that 
\[\lim_{\lambda \to 0} \gamma_\lambda(\sigma,y)=  \left(1-
\E_\sigma[y(R)^N]\right) \frac{y(L)}{y(L)+y(M)} + \E_\sigma[y(R)^N]\left(y(M) +\frac12 y(R)\right).\]
This concludes the proof of the first statement.

 \medskip

 We next assume that $\E_\sigma[N]<+\infty$, and  prove the second part. We will prove that the family of functions $\gamma_\lambda(\sigma,\cdot): y\in \Delta(B) \mapsto \gamma_\lambda(\sigma,y)$ indexed by $\lambda$ is uniformly equicontinuous. Since $\lambda \mapsto \gamma_\lambda (\sigma, y)$ converges pointwise to $\gamma(\sigma,y)$ for each $y$, the Ascoli-Arzela Theorem will then imply that the convergence is uniform.

To prove the uniform equicontinuity claim, it is sufficient to find, for each $\ep>0$, a uniformly equicontinuous family $(g_\lambda)_\lambda$ of functions $g_\lambda: \Delta(B)\to \dR$ such that $\sup_{\lambda,y} \left| \gamma_\lambda(\sigma,y)-g_\lambda(y)\right|< \ep$.

Fix $\ep>0$. Since $\E_\sigma[N]<+\infty$, there exists $T\in \dN$ such that $\prob_\sigma(a_t= \texttt{Top} \mbox{ for some } t\geq T) <\ep$. 

Define $\displaystyle g_\lambda(y):= \E_{\sigma,y}\left[\lambda \sum_{t=1}^T (1-\lambda)^{t-1}r_t\right]  +(1-\lambda)^T \left(\prob_{\sigma,y}(s_T= 1^*)+\prob_{\sigma,y}(s_T= s)\left(y(M) +\frac12 y(R)\right)\right)$.


That is, $g_\lambda(y)$ is the $\lambda$-discounted payoff induced by the strategy that coincides with $\sigma$ until stage $T$ and repeatedly plays \texttt{Bottom} afterward. 
The choice of $T$ guarantees that $\sup_{\lambda,y} \left|  \gamma_\lambda(\sigma,y)-g_\lambda(y)\right|< \ep$. On the other hand, $(\lambda,y)\mapsto g_\lambda(y)$ is a polynomial function, hence the family $(g_\lambda)_\lambda$ is uniformly equicontinuous over the compact set $[0,1]\times \Delta(B)$.

\hfill \halmos
\endproof

Lemma \ref{lemm1} implies Corollary \ref{cor carac} below, which provides conditions on (the moment-generating function of)  $N$ under which $\sigma$ is Blackwell $\ep$-optimal. We note that r.v.'s 
$N$ that satisfy $\E[q^N]= \frac{1}{2-q}$ are geometric random variables over $\dN\cup \{0\}$, with parameter $\frac12$.\medskip

\begin{corollary}\label{cor carac} Let $\sigma$ be a blind strategy.
\begin{description}
    
\item[{\bf C1}] If $\sigma$ is  Blackwell $\ep$-optimal for some $\ep>0$, then 
    \[\left|\E_\sigma[q^N]-\frac{1}{2-q}\right| \leq 2\ep \mbox{ for each } q\in [0,1).\]
\item[{\bf C2}] On the other hand, if  $\E_\sigma[N]<+\infty$ and  $\E_\sigma[q^N]= \frac{1}{2-q}$ for each $q\in [0,1)$, then $\sigma$ is Blackwell $\ep$-optimal for each $\ep>0$.
\end{description}
\end{corollary}

\proof{Proof of Corollary \ref{cor carac}.}
We start with \textbf{C1}. Let $\sigma$ be a Blackwell $\ep$-optimal, blind strategy so that \begin{equation}\label{eq2}
\gamma(\sigma,y)\geq \frac12- \ep \mbox{ for each }y\in \Delta(B).\end{equation} 

Let $q\in [0,1)$ be arbitrary. Let $y_1,y_2\in \Delta(B)$ be defined respectively by $y_1(R)= q$ and $y_1(M)= 1-q$, and by $y_2(R)= q$ and $y_2(L)= 1-q$. Using Equation \eqref{eq1}, one has
\begin{align*}
    \gamma(\sigma,y_1) & = \left(1-\frac{q}{2}\right) \E_\sigma[q^N] \\
    \gamma(\sigma,y_2) & = 1- \E_\sigma[q^N]\left(1-\frac{q}{2}\right).
\end{align*}
Using Equation \eqref{eq2}, the formula for $\gamma(\sigma, y_1)$ implies 
\[\E_\sigma[q^N]\geq \frac{1-2\ep}{2-q}\geq \frac{1}{2-q}-2\ep,\]
and the formula for $\gamma(\sigma,y_2)$ yields 
\[\E_\sigma[q^N]\leq \frac{1+2\ep}{2-q}\leq  \frac{1}{2-q}+2\ep.\] 
These two inequalities imply \textbf{C1}.
\medskip

We now prove {\bf C2}. Assume that $\E_\sigma[N]<+\infty$ and that $\E_\sigma[q^N]= \frac{1}{2-q}$ for each $q\in [0,1)$. By Lemma \ref{lemm1}, $\lim_{\lambda \rightarrow 0} \gamma_\lambda(\sigma,y)= \frac12$ for each $y\in \Delta(B)$, and the convergence is uniform with respect to $y$. This implies that for each $\ep>0$, there is $\lambda_0>0$ such that 
\[\gamma_\lambda(\sigma,y)\geq \frac12 -\ep,\mbox{ for each } \lambda <\lambda_0 \mbox{ and } y\in \Delta(B),\]
hence $\sigma$ is Blackwell $\ep$-optimal for each $\ep>0$.
\hfill \halmos
\endproof
\medskip

It follows from Corollary \ref{cor carac} that any blind strategy $\sigma$ such that $N$ follows a geometric distribution over  $\dN\cup\{0\}$ with parameter $\frac12$ is  Blackwell $\ep$-optimal for each $\ep>0$.    
There are many such strategies. A particularly simple one is the strategy $\sigma_*$ that assigns equal probabilities to \texttt{Top} and \texttt{Bottom} as long as P1 has played \texttt{Top} in all previous rounds, and that repeats \texttt{Bottom} forever once \texttt{Bottom} has been played once. The strategy $\sigma_*$ is implemented by the following two-state autonomous automaton: (i) the state space is $K= A$, (ii) the initial internal state is drawn according to the uniform distribution $\mu_0$ over $K$, (iii) the action function $f:K \rightarrow A$ is $f(k)= k$, (iv) when in state $k= \texttt{Top}$, the next state is drawn according to $\mu_0$; on the other hand, the state $k= \texttt{Bottom}$ is absorbing.
We stress that, while the states are labeled after actions for transparency, the transitions only involve the internal state of the automaton.

\usetikzlibrary {arrows.meta,automata,positioning}
\medskip
\begin{center}
    \begin{tikzpicture}
  [shorten >=1pt,node distance=2cm,on grid,>={Stealth[round]},initial text=,
   every state/.style={draw=blue!50,very thick},
   accepting/.style=accepting by arrow]

  \node[state, draw= none]  (q_T) {};
  \node[state, draw= none, below of=q_T]  (q_B) {};
  \node[state]          (T) [right of =q_T] {$T$};
  \node[state]          (B) [right of =q_B] {$B$};

  \path[->] (q_T) edge              node [above left]  {1/2} (T)
            (q_B)  edge              node [below left]  {1/2} (B)
            (T) edge              node [ right] {1/2} (B)
                  edge [loop above] node               {1/2} ()
            (B) edge [loop below] node               {1} ();
\end{tikzpicture}

The automaton that implements $\sigma_*$
\medskip
\end{center}

More generally, let for each $n\geq 0$ a sequence $\vec{a}^n= (a_t^n)_{t \in \dN}$ of actions with exactly $n$ occurrences of \texttt{Top}. Then the mixed strategy $\sigma$ that randomly selects one of these sequences  (with probabilities $\frac{1}{2^{n+1}}$, $n\in \dN\in \{0\}$) prior to  the beginning of the game, then follows it throughout the game, is also Blackwell $\ep$-optimal for each $\ep>0$.  

Finally, any mixture of such blind strategies remains Blackwell $\ep$-optimal for each $\ep>0$.

\subsubsection{On Markovian strategies in the Modified Big Match}\label{proof second}

We here prove that P1 has no Markovian Blackwell $\ep$-optimal strategy in the modified Big Match. This follows from Lemma \ref{lemm2} below and completes the proof of  Proposition \ref{prop MBM}. 
\begin{lemma}\label{lemm2}
 There exists $\ep>0$ such that the following holds: for each Markovian strategy $\sigma$, there exists $y\in \Delta(B)$ such that $\gamma(\sigma,y)\leq \frac12 -\ep$.
 \end{lemma}

We fix $c\in \dR$ such that $\ln 2 < c < 2$, and $q\in (0,1)$
 such that $\phi_c(q)>\frac12$, where $\phi_x(q):=e^{-qx}\left(\frac12 +q\right) - q^2 x^2$. The existence of such a $q \in (0,1)$ follows from the fact that $\phi_c(0)= \frac12$ and $\phi'_c(0)= 1-\frac{c}{2}>0$. We choose $\ep>0$ such that  $\ep<\min\left(\phi(q)-\frac12,\frac12 - e^{-c}\right)$. We prove that the conclusion of Lemma \ref{lemm2} holds for this choice of $\ep$.
 \medskip

 Let $\sigma$ be an arbitrary Markovian strategy. We set $X_t=1$ if $a_t= \texttt{Top}$ and $X_t=0$ otherwise, so that $\displaystyle N=\sum_{t=1}^{+\infty} X_t$. Since $\sigma$ is Markovian, the variables $(X_t)_{t \in \dN}$ are independent Bernoulli variables with parameters $p_t:=\prob_\sigma(X_t=1)$. 
\medskip

We will denote by $M$ the stationary strategy of P2 that always plays \texttt{Middle}. Given $(\sigma, M)$, the game moves to $0^*$ as soon as P1 plays \texttt{Top}, and otherwise remains in the non-absorbing state $s$, with a payoff of 1 in every stage. Therefore, the realized discounted payoff $\bar r_\lambda$
satisfies $\lim_{\lambda \rightarrow 0} \bar r_\lambda= 0$ if $N>0$ and 
$\lim_{\lambda \rightarrow 0} \bar r_\lambda= 1$ if $N=0$. By dominated convergence, 
$\gamma(\sigma,M) = \prob_\sigma(N=0)$.
 If $\displaystyle\sum_{t=1}^{+\infty} p_t=+\infty$, the Borel-Cantelli lemma implies that  $N=+\infty$ $\prob_\sigma$-a.s., so that $\gamma(\sigma,M)= 0$.
We now assume that $\displaystyle \sum_{t=1}^{+\infty} p_t<+\infty$ and discuss two cases.
\medskip

\noindent
\textbf{Case 1}: $\displaystyle \sum_{t=1}^{+\infty} p_t \geq c$. In this case,  P1 is more likely to never play \texttt{Top}. Indeed, 
\[\prob_\sigma(N=0)= \prod_{t=1}^{+\infty} (1-p_t) \leq e^{-\sum_{t=1}^{+\infty} p_t} \leq e^{-c}< \frac12 -\ep.\]
Hence,  
$\gamma(\sigma,M) = \prob_\sigma(N=0)< \frac12 -\ep.$

\medskip\noindent
\textbf{Case 2}: $\displaystyle \sum_{t=1}^{+\infty} p_t \leq c$. We will use the result below, which is a weaker version of Le Cam's inequality \cite{le1960approximation}.
 
 \begin{lemma}\label{lemm lecam}
 Let $Z_t$, $t\geq 1$ be independent Bernoulli variables with parameters $q_t$, such that $\displaystyle \sum_{t\geq 1} q_t<+\infty$. Set $N_Z:=\sum_{t\geq 1} Z_t$ and let $Z\sim \calP(\sum_{t\geq 1} q_t)$ be a Poisson variable with parameter $\sum_{t\geq 1}q_t$. Then 
 \[\left|\prob(N_Z= k)-\prob(Z=k)\right|\leq \sum_{t\geq 1} q_t^2, \mbox{ for each } k\in \dN.\]
 \end{lemma}
 We will show that $\gamma(\sigma,y_q)<\frac12-\ep$, where $y_q=(0,q,1-q)\in \Delta(B)$ assigns probabilities $q$ and $1-q$ to \texttt{Middle} and \texttt{Right} (recall that $q \in (0,1)$ is s.t. $\phi_{c}(q) >\frac12+\ep$). Set $Y_t:= 1_{b_t=\texttt{Middle}}$ and $Z_t=X_tY_t$, so that $(Z_t)_{t \in \dN}$ are independent Bernoulli variables with parameters $p_tq$. 

 Given $(\sigma,y_q)$, the game ends in $S^*$ if and only $(a_t,b_t)= (\texttt{Top}, \texttt{Middle})$ for some $t \in \dN$. Hence,  $\tau<+\infty$ coincides with the event $N_Z>0$. In that case, the game ends in $1^*$ and  $\lim \bar r_\lambda= 1$. 
 On the other hand, 
 the empirical distribution of $(b_t)_{t \in \dN}$ converges to $y_q$, hence $\lim_{\lambda \rightarrow 0}\bar r_\lambda=\frac12 y_q(R)= \frac12 (1-q)$, $\prob_{\sigma, y_q}$-a.s. on the event $\tau=+\infty$. By dominated convergence, it follows that 
\[\gamma(\sigma,y_q)= \prob_{\sigma,y_q}\left(N_Z>0\right) +\frac12 (1-q) \prob_{\sigma,y_q}\left(N_Z=0\right).\]
By Lemma \ref{lemm lecam}, and with $Z\sim \calP\left(\sum_{t\geq 1}qp_t\right)$, one has 
\begin{align*}
    \gamma(\sigma,y_q) &\leq \prob_{\sigma,y_q}\left(Z>0\right) +\frac12 (1-q) \prob_{\sigma,y_q}\left(Z=0\right)+  \sum_{t\geq 1} (qp_t)^2\\
    &=1-e^{-q\sum_t p_t} +\frac12(1-q) e^{-q\sum_{t\geq 1} p_t}+ q^2 \sum_{t\geq 1} p_t^2 \\
    &\leq 1- \phi_{\sum_{t\geq 1}p_t}(q) \\
    &\leq  1-\phi_c(q) \\
    &\leq \frac12 -\ep,
\end{align*}
where the first inequality follows from Lemma \ref{lemm lecam}, the equality uses $Z\sim \calP(\sum_{t\geq 1} q_t)$, the second inequality follows from $\sum_{t\geq 1}p_t^2\leq \left(\sum_{t\geq 1} p_t\right)^2$, the third inequality from the assumption $\sum_{t\geq 1}p_t\leq c$ and from the fact that $x\mapsto \phi_x(q)$ is non-increasing, and the last inequality from the inequality $\phi_c(q)>\frac12 +\ep$. 

\subsection{Blackwell $\ep$-optimal strategies in product absorbing games}\label{sec GBM}
We now prove the second statement of Theorem \ref{th main}, namely, the existence of Blackwell $\ep$-optimal,  autonomous automata of size $2$ in product absorbing games.

We let $\Gamma$ be a product absorbing game.
We recall that  $p^*(a,b):=p(S^*\mid a,b)$ for $(a,b)\in A\times B$, and we set $g^*(a,b):=\sum_{s^*\in S^*} p(s^*\mid a,b) r(s^*)$.
We build on the following formula for $\lim_{\lambda \rightarrow 0} v_\lambda$ from \cite{cardaliaguet2012continuous}, first proven in \cite{laraki2010explicit} in a  different form: 
\begin{equation}\label{limit value}
\lim_{\lambda \rightarrow 0} v_\lambda= \mathrm{val}_{(x,\alpha),(y,\beta)\in (\Delta(A)\times \dR_+^A)\times (\Delta(B)\times \dR_+^B)} \frac{r(x,y)+ g^*(\alpha,y)+g^*(x,\beta)}{1+p^*(\alpha,y)+p^*(x,\beta)},
\end{equation}
where $p^*$ and $g^*$ denote the bilinear extensions of $p^*$  and $g^*$ to $\dR^A\times \dR^B$. That is, $\lim_{\lambda \rightarrow 0} v_\lambda$ is the value of a one-shot game, in which the action set of P1 consists of pairs $(x,\alpha) \in \Delta(A) \times \dR_{+}^{A}$, the action set of P2 consists of pairs $(y,\beta) \in \Delta(B) \times \dR_{+}^{B}$ and the payoff is as specified on the right-hand side of \eqref{limit value}.
\medskip

Let $\ep>0$ be given, and set $\eta=\frac{\ep}{2}$.
Let $(x,\alpha)\in  \Delta(A)\times \dR^A$ be an $\eta$-optimal strategy of P1 in the right-hand-side of \eqref{limit value}. That is, the pair $(x,\alpha)$ is such that 
\begin{equation}\label{opt}
 \frac{r(x,y)+ g^*(\alpha,y)+g^*(x,\beta)}{1+p^*(\alpha,y)+p^*(x,\beta)} \geq \lim_{\lambda \rightarrow 0} v_\lambda - \eta, \forall \; (y,\beta)\in \Delta(B)\times \dR_{+}^{B}.
\end{equation}
 
 We next define a strategy $\sigma_\ep$.
 If $\alpha(a)= 0$ for each $a\in A$, or if $x(a^*)>0$ for some $a^*\in A^*$, we let $\sigma_\ep:= x$ be the stationary strategy that repeats $x$ in state $s$.

 If instead $\alpha\neq 0_{\dR^A}$ and $x(A^*)=0$, $\sigma_\ep$ is defined as follows. Set $\bar \alpha= \sum_{a \in A} \alpha(a)$ and $\displaystyle x_\alpha:= \frac{\alpha}{\bar \alpha} \in \Delta(A)$. Intuitively, $\bar \alpha$ is related to the intensity with which $x_\alpha$ is played in an optimal strategy $x_\lambda$ in the $\lambda$-discounted game.
Let $D$  be a geometric random variable over $\dN\cup \{0\}$ with parameter $\delta: = \frac{1}{1+\bar \alpha}$: $\prob(D= k)= \delta(1-\delta)^k$ for each $k\geq 0$. 
The strategy $\sigma_\ep$ plays  $x_\alpha$ in all stages $t=1,2,\ldots, D$, and $x$ in later stages.\footnote{We define $\sigma_\ep$ as a mixture over behavior strategies. By Kuhn Theorem, it could equivalently be described as a mixed strategy or as a behavior strategy.}

We note that $\sigma_\ep$ can be implemented by a two-state autonomous automaton, as follows. Let the set of internal states be  $K=\{k_\alpha,k^*\}$, and the action function as $f(k_\alpha)= x_\alpha$ and $f(k^*)= x$. The initial distribution assigns probability $\delta$ to $k^*$. The internal state $k^*$ is absorbing, and the automaton moves from $k_\alpha$ to $k^*$ with probability $\delta$. In this way, the number of stages before the automaton enters the internal state $k^*$ follows the same distribution as $D$, and the automaton chooses $x_\alpha$ as long as the internal state is $k_\alpha$. 

\medskip
We prove below that $\sigma_\ep$ is a Blackwell $\ep$-optimal strategy.
For $\lambda>0$, we denote by $y_\lambda$ a solution to the minimization problem $\min_{y\in \Delta(B)} \gamma_\lambda(\sigma_\ep,y)$, that is, $y_\lambda$ is a best response of P2 to $\sigma_\ep$ in the $\lambda$-discounted game, when restricted to stationary strategies. The existence of $y$ follows from the continuity of the map $y\in \Delta(B)\mapsto \gamma_\lambda(\sigma_\ep,y)$.  Since $\sigma_\ep$ need not be stationary, $y_\lambda$ need not be an unconditional best response of P2. We prove the result below, which implies that $\sigma_\ep$ is Blackwell $\ep-$optimal.

\begin{proposition}\label{prop suff}
    One has  $\lim_{\lambda \rightarrow 0} \gamma_\lambda(\sigma_\ep,y_\lambda)\geq \lim_{\lambda \rightarrow 0} v_\lambda - \eta$.
\end{proposition}

The rest of this section is devoted to the proof of Proposition \ref{prop suff}.
We first prove the result in the case where $\sigma_\ep$ is stationary. \medskip

\paragraph{The case $\sigma_\ep$ stationary.}

This is the case where $x(A^*)>0$ and/or $\alpha = 0$. We first derive a few implications of (\ref{opt}).
\begin{description}
    \item[\bf{P1}]: $r(x,y)\geq  \lim_{\lambda \rightarrow 0} v_\lambda-\eta$ for every $y\in \Delta(B)$ such that $y(B^*)=0$.
    \item[\bf{P2}]: If $\alpha = 0$, one has $r(x,y)\geq  \lim_{\lambda \rightarrow 0} v_\lambda-\eta$ for every $y\in \Delta(B)$.
    \item[\bf{P3}]: If $x(A^*)>0$, one has $\displaystyle \frac{g^*(x,y)}{p^*(x,y)}\geq \lim_{\lambda \rightarrow 0} v_\lambda-\eta$ whenever   $y(B^*)>0$.
\end{description}

Indeed, if $y(B^*)=0$, then $p^*(\alpha,y)=g^*(\alpha,y)=0$. Setting $\beta=0$ in (\ref{opt}) yields \textbf{P1}. Next, if $\alpha=0$,  setting $\beta=0$ in (\ref{opt}) yields $r(x,y)\geq  \lim_{\lambda \rightarrow 0} v_\lambda-\eta$, which is \textbf{P2}.
Finally, if $x(A^*)>0$ and $y(B^*)>0$, then (\ref{opt}) implies 
\[ \frac{r(x,y)+ g^*(\alpha,y)+n \times g^*(x,y)}{1+p^*(\alpha,y)+n \times p^*(x,y)} \geq \lim_{\lambda \rightarrow 0} v_\lambda - \eta, \mbox{ for each }n\in \dN.\]
\textbf{P3} follows when letting $n\to +\infty$, since $p^*(x,y)>0$.
\medskip

Note now that the identity $\gamma_\lambda(x,y_\lambda)= \lambda r(x,y_\lambda) +(1-\lambda )
\left(g^*(x,y_\lambda) +(1-p^*(x,y_\lambda))r(x,y_\lambda)\right)$ implies 
\begin{equation}\label{eq pay}\gamma_\lambda(x,y_\lambda)= \frac{\lambda r(x,y_\lambda)+ (1-\lambda )g^*(x,y_\lambda)}{1-(1-\lambda)(1-p^*(x,y_\lambda))}.\end{equation}

W.l.o.g., we may assume that $\bar y:=\lim_{\lambda \to 0}y_\lambda$ exists. Assume first that $\bar y(B^*)>0$. If $x(A^*)>0$ then taking the limit $\lambda \to 0$ in (\ref{eq pay}) yields $\displaystyle \lim_{\lambda \rightarrow 0} \gamma_\lambda(x,y_\lambda)= \frac{g^*(x,\bar y)}{p^*(x,\bar y)}$, which is at least $\lim_{\lambda \rightarrow 0} v_\lambda-\eta$ by \textbf{P3}. If $x(A^*)=0$, then it must be that $\alpha=0$, since we assume here $\sigma_{\ep}$ to be stationary. Since $g^*(x,y_\lambda)=p^*(x,y_\lambda)=0)$, one has $\gamma_\lambda(x,y_\lambda)= r(x,y_\lambda)$ by (\ref{eq pay}), hence  $\displaystyle \lim_{\lambda \rightarrow 0} \gamma_\lambda(x,y_\lambda)=r(x,\bar y)$  which is at least $\lim_{\lambda \rightarrow 0} v_\lambda-\eta$ by \textbf{P2}.

If instead $\bar y(B^*)=0$, we may decompose $y_\lambda = y_\lambda^a+y_\lambda^n$, where $y_\lambda^a$ and $y_\lambda^n$ are the restrictions of $y_\lambda$ to $B^*$ and $B\setminus B^*$ respectively. In particular, $\lim_{\lambda \rightarrow 0} y_\lambda^a=0$ and $\lim_{\lambda \rightarrow 0} y_\lambda^n= \bar y$. Substituting into (\ref{eq pay}), and observing that $p^*(x,y_\lambda)= p^*(x,y_\lambda^a)+p^*(x,y_\lambda^n)=p^*(x,y_\lambda^a)$, one obtains 
\begin{equation}\label{eq pay2}
   \gamma_\lambda(x,y_\lambda)= \frac{\lambda r(x,\bar y)+  g^*(x,y^a_\lambda)}{\lambda+p^*(x,y^a_\lambda)} +o(1).
\end{equation}
By \textbf{P1}, $ r(x,\bar y)\geq \lim_{\lambda \rightarrow 0} v_\lambda-\eta$. By   \textbf{P3}, $\displaystyle\frac{g^*(x,y^a_\lambda)}{p^*(x,y^a_\lambda)}\geq \lim_{\lambda \rightarrow 0} v_\lambda-\eta$ if $x(A^*)>0$ and $g^*(x,y^a_\lambda)= p^*(x,y^a_\lambda)=0$ if $x(A^*)=0$. Either way, this yields $\gamma_\lambda(x,y_\lambda)\geq \lim_{\lambda \rightarrow 0} v_\lambda-\eta -o(1)$ using (\ref{eq pay2}),  hence $\lim_{\lambda \rightarrow 0} \gamma_\lambda(x,y_\lambda) \geq \lim_{\lambda \rightarrow 0} v_\lambda-\eta$.

\paragraph{The case $\sigma_\ep$ non-stationary.}

This is the case where $x(A^*)=0$ and $\alpha\neq 0$. The conclusion  will follow from 
Lemma \ref{lem limpay} below.

\begin{lemma}\label{lem limpay}
    One has $\gamma_\lambda(\sigma_\ep,y_\lambda)= \frac{r(x,y_\lambda)+g^*(\alpha,y_\lambda)}{1+p^*(\alpha,y_\lambda)}+o(1).$
\end{lemma}

We recall that $\gamma_\lambda(\sigma_\ep,y_\lambda)=\E_{\sigma_\ep,y_\lambda}[\bar r_\lambda]$ where the realized discounted payoff is equal to 
\[\bar r_\lambda= \sum_{t=1}^{\tau-1} \lambda(1-\lambda)^{t-1} r_t + (1-\lambda)^\tau r^*(s^*_\tau)= \sum_{t=1}^{+\infty}\lambda(1-\lambda)^{t-1} 1_{\tau >t} r_t + (1-\lambda)^\tau r^*(s^*_\tau).\]

The proof of Lemma \ref{lem limpay} relies  on Claims \ref{claim1} through \ref{claim3} below.\footnote{The notation $o(1)$ stands for any expression that converges to zero as $\lambda \to 0$.}
\begin{claim}\label{claim1}
    One has 
    \[\E_{\sigma_\ep,y_\lambda}\left[ \sum_{t=1}^{+\infty}\lambda(1-\lambda)^{t-1} 1_{\tau >t} r_t\right]= r(x,y_\lambda) \prob(\tau> D+1) + o(1).\]
\end{claim}

\proof{Proof.}
Note first that 
\begin{align*}
     \E_{\sigma_\ep,y_\lambda}\left[ \sum_{t=1}^{D}\lambda(1-\lambda)^{t-1} 1_{\tau >t} \left|r_t\right|\right] & \leq \|r\|_\infty\E_{\sigma_\ep,y_\lambda}\left[ \sum_{t=1}^{D}\lambda(1-\lambda)^{t-1}\right] \\
     & = \|r\|_\infty \left(1-\E_{\sigma_\ep,y_\lambda}\left[(1-\lambda)^D\right]\right). 
\end{align*}
Since $D<+\infty$ $\prob_{\sigma_\ep,y_\lambda}$-a.s., one has $\lim_{\lambda \rightarrow 0} \E_{\sigma_\ep,y_\lambda}\left[(1-\lambda)^D\right]= 1$, hence 
the left-hand side $\displaystyle \E_{\sigma_\ep,y_\lambda}\left[ \sum_{t=1}^{D}\lambda(1-\lambda)^{t-1} 1_{\tau >t} \left|r_t\right|\right]$ converges to 0.

Since $\sigma_\ep$ coincides with $x$ beyond stage $D$ and since $x(A^*)=0$, one has  either $\tau\leq D+1$ or $\tau= +\infty$ and 
\begin{align*}
    \E_{\sigma_\ep,y_\lambda}\left[ \sum_{t=D+1}^{+\infty}\lambda(1-\lambda)^{t-1} 1_{\tau >t} r_t\right] & = r(x,y_\lambda )
\E_{\sigma_\ep,y_\lambda}\left[1_{\tau>D+1} \sum_{t=D+1}^{+\infty}\lambda(1-\lambda)^{t-1} \right] \\
& = r(x,y_\lambda )\E_{\sigma_\ep,y_\lambda}\left[1_{\tau>D+1} (1-\lambda)^D\right].
\end{align*}
Since $\E_{\sigma_\ep,y_\lambda}\left[1_{\tau>D+1} (1-\lambda)^D\right]$ converges to $\prob(\tau>D+1)$, the latter identity implies
\[\E_{\sigma_\ep,y_\lambda}\left[ \sum_{t=D+1}^{+\infty}\lambda(1-\lambda)^{t-1} 1_{\tau >t} r_t\right]=r(x,y_\lambda)\prob(\tau>D+1) +o(1),\] 
as desired
\hfill \halmos
\endproof

\begin{claim}\label{claim2}
    One has 
        \[\E_{\sigma_\ep,y_\lambda}\left[ (1-\lambda)^\tau r^*(s^*_\tau)1_{\tau <+\infty}  \right]= 
    \frac{g^*(x_\alpha,y_\lambda)}{p^*(x_\alpha,y_\lambda)} \prob(\tau \leq D+1)+o(1).\]
\end{claim}
\proof{Proof.} Since either $\tau= +\infty$ or $\tau\leq D+1$ and since $(1-\lambda)^\tau $ converges to 1 if $\tau<D+1$, one has 
\[\displaystyle \E_{\sigma_\ep,y_\lambda}\left[ (1-\lambda)^\tau r^*(s^*_\tau)1_{\tau <+\infty}  \right]=\E_{\sigma_\ep,y_\lambda}\left[  r^*(s^*_\tau)1_{\tau <+\infty}  \right] +o(1).\] For every stage $t\geq 1$, the conditional terminal payoff, conditional on $\tau= t$, is $\displaystyle \frac{g^*(x_\alpha,y_\lambda)}{p^*(x_\alpha,y_\lambda)}$. This implies that $\displaystyle\E_{\sigma_\ep,y_\lambda}\left[  r^*(s^*_\tau)1_{\tau <+\infty}  \right]=  \frac{g^*(x_\alpha,y_\lambda)}{p^*(x_\alpha,y_\lambda)} \prob(\tau \leq D+1)$.
\hfill \halmos
\endproof

\begin{claim}\label{claim3}
    One has 
    \[\prob_{\sigma_\ep, y_\lambda}\left(\tau >D+1\right)= \frac{1}{1+ p^*(\alpha,y_\lambda)}.\]
\end{claim}

\proof{Proof.}
One has $\prob_{\sigma_\ep, y_\lambda}\left(\tau >D+1\right)= \displaystyle \E_{\sigma_\ep,y_\lambda}[(1-p^*(x_\alpha,y_\lambda))^D]$. 
Since $D$ follows a geometric distribution over $\dN\cup \{0\}$ with parameter $\delta$, this expectation is given by 
\[\E_{\sigma_\ep,y_\lambda}[(1-p^*(x_\alpha,y_\lambda))^D]= \frac{\delta}{1-(1-\delta)(1-p^*(x_\alpha,y_\lambda))}= \frac{1}{1+\bar \alpha p^*(x_\alpha,y_\lambda)}=\frac{1}{1+ p^*(\alpha,y_\lambda)},\]
where the first equality follows from the expression of the moment generating function of $D$, the second equality obtains when replacing $\delta= \frac{1}{1+\bar \alpha}$, and the third one follows since $\bar \alpha p^*(x_\alpha,y_\lambda)= p^*(\bar \alpha x_\alpha,y_\lambda)= p^*(\alpha,y_\lambda)$.
\hfill \halmos
\endproof

\medskip

We are now in a position to prove Lemma \ref{lem limpay}.\medskip

\proof{Proof of Lemma \ref{lem limpay}.}
    Claims \ref{claim1} and \ref{claim2} yield
    \begin{equation}\label{eq4}
        \gamma_\lambda(\sigma_\ep,y_\lambda)= \prob(\tau>D+1)r(x,y_\lambda) 
+ \frac{g^*(x_\alpha, y_\lambda)}{p^*(x_\alpha, y_\lambda)} \prob(\tau \leq D+1) +o(1).
    \end{equation}
Claim \ref{claim3} yields
\[\frac{g^*(x_\alpha, y_\lambda)}{p^*(x_\alpha, y_\lambda)} \prob(\tau \leq D+1)=g^*(x_\alpha, y_\lambda)\times \frac{p^*(\alpha, y_\lambda)}{p^*(x_\alpha, y_\lambda)} \times \frac{1}{1+p^*(\alpha,y_\lambda)}= \frac{g^*(\alpha,y_\lambda)}{1+p^*(\alpha,y_\lambda)}. \]
The result follows when substituting into (\ref{eq4}).
\hfill \halmos
\endproof
\section{Blackwell optimality for absorbing games}\label{sec general}
In this section, we show that the positive results obtained for product absorbing games do not extend to absorbing games. 
We consider the absorbing game described by Table 3 below, and prove that P1 has no blind, Blackwell $\ep$-optimal strategy for $\ep$ small enough, thus proving Proposition \ref{prop counter}.
\[ \begin{array}{|c|c|c|}
   \hline
    1^* & 0^* & \frac12^*\\
    \hline
    0 & 1 & \frac12^*\\
    \hline
    \end{array}
    \]
\begin{center}
Table 3
\end{center}

For the same reasons as the ones that apply to the modified Big Match of Section \ref{sec modified}, the discounted value of this game is $v_\lambda = \frac12$, for each $\lambda >0$. We note that, as in Section \ref{sec GBM}, the limit $\gamma(\sigma,y)= \lim_{\lambda \to 0} \gamma_\lambda(\sigma,y)$ exists for each blind strategy $\sigma$ and $y\in \Delta(B)$.
Hence Proposition \ref{prop counter}  readily follows from Lemma \ref{ex abs} below.

\begin{lemma}\label{ex abs}
For every blind strategy $\sigma$ of P1, one has $\inf_{y\in \Delta(B)}  \gamma(\sigma,y)\leq \frac13$.
\end{lemma}

\proof{Proof.}
    Let $\sigma$ be an arbitrary blind strategy of P1, and set $N:=\#\{t\geq 1: a_t= \texttt{Top}\}$. Since $\sigma$ is blind, the distribution of $N$ does not depend on $y$. As above, we denote by $L$ and $M$ the two pure stationary strategies of P2 that play respectively the \texttt{Left} and \texttt{Middle} columns in every stage. We denote by 
    $T_1:=\inf\{t\geq 1: a_t= \texttt{Top}\}$ the first stage where P1 chooses the top row. We discuss three cases.

    \paragraph{Case 1: $\prob_\sigma(N>0)\geq \frac23$.}

Given the profile $(\sigma,M)$, the realized discounted payoff $\bar r_\lambda$ converges to 0 on the event $N>0$ (as $\lambda \to 0$), and is equal to 1 if $N=0$. Hence 
$\gamma(\sigma,y)= \prob(N=0)\leq \frac13$.

    \paragraph{Case 2: $\prob_\sigma(N>0)\leq \frac13$.}
Given the profile $(\sigma,L)$, the realized discounted payoff $\bar r_\lambda$ converges to 1 on the event $N>0$, and is equal to 0 if $N=0$. Hence 
$\gamma(\sigma,y)= 1-\prob(N=0)\leq \frac13$.

    \paragraph{Case 3: $\prob_\sigma(N>0)\in \left(\frac13, \frac23\right)$.}

Let $\ep>0$ be arbitrary. Since the event $\{N>0\}$ is the increasing union of the sets $\{T_1\leq T\}$, $T\in \dN$, there is $\bar T\in \dN$ such that $\prob_\sigma(T_1\leq  \bar T)\geq \prob(N>0)-\ep\geq \frac13 -\ep$. Set $\eta= \frac{\ep}{\bar T}$ and let $y_\eta\in \Delta(B)$ be the stationary strategy that plays $\texttt{Middle}$ with probability $1-\eta$, and $\texttt{Right}$ with probability $\eta$. 

By the union bound, the probability that P2 plays at least once \texttt{Middle} in the first $\bar T$ stages is at most $\ep$. Denoting $T_2:=\inf\{t: b_t=\texttt{Right}\}$, this implies that $\prob_{\sigma,y_\eta}\left(T_1<T_2\right)\geq \prob_{\sigma,y_\eta}\left(T_1\leq \bar T<T_2\right)\geq \frac13 -2\ep$. Note now that,  $\prob_{\sigma,y_\eta}$-a.s, the game ends in the absorbing state $0^*$ if $T_1<T_2$, and  in the absorbing state $\frac12^*$ if $T_1\geq T_2$. This implies
\[\gamma(\sigma,y_\eta)= \frac12 \times \prob_{\sigma,\eta}(T_1\geq T_2)\leq \frac12\times \left(\frac23 +2\ep\right)= \frac13 +\ep.\]
Since $\ep>0$ is arbitrary, the result follows.
\hfill \halmos
\endproof

\section{Implications for robust MDPs}\label{sec RMDPs}
In this section we briefly comment on the implications of our results for robust MDPs.
An $s$-rectangular robust MDP instance~\cite{wiesemann2013robust} is a tuple $\M = (S, A,r, \U)$ where $S$ and $A$ are the state and action sets, $r: S\times A\times S\rightarrow \dR$ is the reward function of the decision maker and $\U = \times _s \U_s $ is the uncertainty set (assumed to be convex), i.e. the set of all conceivable transition probabilities, with $\U_s \subseteq \Delta(S)^A$. 
The decision maker chooses a \emph{policy}, which maps the sequence of past and current states and (own) actions into $\Delta(A)$, while nature chooses  the transitions probabilities $\prob\in \U$
associated with the various actions in different states. 

If $S$ and $A$ are finite, and if $\U_s$ is a polyhedral subset of $\Delta(S)^A$ for each $s$, such an instance can be described as a (zero-sum) game in which the finite action set $B_s$ of P2 in state $s$ is identified with $\ext(U)$ the set of extreme points of $\U_s$. The reward $\tilde r(s,a,b)$ in this auxiliary SG is obtained as follows. Given $s$, each action $b$ is a collection $(p(\cdot \mid s,a))_{a\in A}$ of distributions over $S$, indexed by $A$. We simply set $\tilde r(s,a,b):=\E_{s'\sim p(\cdot\mid s,a)}[r(s,a,s')]$. Conversely, any finite SG induces a (polyhedral) $s$-rectangular robust MDP instance, where stationary mixed strategies of the second-player correspond to the adversary choosing probability transitions in the polyhedral uncertainty set, i.e., to the adversary choosing probability distributions over the set of extreme points of the polyhedral uncertainty set. We refer to \cite{iyengar2005robust,chatterjee2023solving,grand2023beyond} for more discussion on the connection between stochastic games and robust MDPs. 

Our Theorem \ref{th main} implies that there exists a class of s-rectangular robust MDP instances where the decision maker has no {\em Markovian} Blackwell $\ep$-optimal policy (for $\ep$ small), yet has a Blackwell $\ep$-optimal policy that can be implemented by an autonomous automaton. This class of robust MDP instances corresponds to the instances constructed from product absorbing stochastic games.
Historically, the study of the properties of optimal policies in robust MDPs has focused solely on distinguishing between stationary, Markovian, and history-dependent policies, the last two being considered difficult to implement. Yet, our results show that for certain classes of robust MDP instances, even if Blackwell $\ep$-optimal policies may not be Markovian,  there always exists a Blackwell $\ep$-optimal that is history-dependent but exhibits a very simple structure (namely, an autonomous automaton of size $2$). To the best of our knowledge, this result is the first to go beyond the distinction ``stationary vs. Markovian vs. history-dependent" policies for robust MDPs and to point to other simple simple subclasses of policies of interest (autonomous automaton). 

Our Proposition \ref{prop counter} implies, on the other hand, that in general, Blackwell $\ep$-optimal policies must depend on the sequence of past states, past chosen actions of the DM, and the past actions chosen by the adversary, that is, on the sequence of extreme points of $\U$ sampled by the stationary adversary over time from the chosen distribution over $\ext(\U)$. This answers a question raised in \cite{grand2023beyond} about the existence and properties of Blackwell $\ep$-optimal policies for s-rectangular robust MDPs.

As a final remark, we note that the literature on SGs traditionally assumes that reward functions map $S\times A \times B$ to $\dR$ and that the literature on (robust) MDPS assumes that reward functions map $S\times A\times S$ to $\dR$. While these two formalisms are `equivalent', the class of (product) absorbing games does not map to a simple class of robust MDPs (since we need to duplicate some states to obtain a robust MDP reformulation from an instance of absorbing stochastic game). We leave the analysis of other specific classes of robust MDPs and other classes of policies for future research.

\bibliography{ref}

\end{document}